\documentclass[
 amsmath,amssymb,
 reprint,%
]
{revtex4-2}
\pdfoutput=1 

\usepackage{graphicx}
\usepackage{dcolumn}
\usepackage{bm}
\usepackage{color}
\usepackage{ulem}

\newcommand{\B}{{\cal B}}

\begin{document}

\preprint{AIP/123-QED}

\title{Topological Diagrams and Hadronic Weak Decays of Charmed Baryons}

\author{Huiling Zhong}
\author{Fanrong Xu}
\email{fanrongxu@jnu.edu.cn}
 \affiliation{Department of Physics, College of Physics $\&$ Optoelectronic Engineering, Jinan University, Guangzhou 510632, P.R. China}

\author{Hai-Yang Cheng}
\affiliation{Institute of Physics, Academia Sinica, Taipei, Taiwan 11529, Republic of China}

\date{\today}

\begin{abstract}
Inspired by the recent BESIII measurement of the decay asymmetry and the phase shift between $S$- and $P$-wave amplitudes in the decay $\Lambda_c^+\to \Xi^0K^+$, we perform a global fit to the experimental data of charmed baryon decays based on the topological diagrammatic approach (TDA) which has the advantage that it is more intuitive and easier to implement model calculations. The measured branching fractions and decay asymmetries are well accommodated in TDA except for three modes, in particular, the predicted $\B(\Xi_c^0\to \Xi^-\pi^+)=(2.83\pm0.10)\%$ is larger than its current value. 
The predicted magnitudes of $S$- and $P$-wave amplitudes and their phase shifts are presented for measured and yet-measured modes which can be tested in forthcoming experiments.
\end{abstract}

\maketitle

\noindent{\it  1. Introduction---} 
Both experimental and theoretical progresses in the study of hadronic decays of charmed baryons were very slow for a long time. The situation was reversed since 2014 as there were several major breakthroughs in 
charmed-baryon experiments in regard to the weak decays of $\Lambda_c^+$ and $\Xi_c^{+,0}$ (for a review, see Ref. 
\cite{Cheng:2021qpd}). Consider the decay $\B_i\to\B_f+P$ with $P$ being a pseudoscalar meson and $\B_i$, $\B_f$ the initial- and final-state baryon, respectively, its general decay amplitude reads 
\begin{eqnarray}
\label{eq:A&B}
M(\B_i\to \B_f+P)=i\bar u_f(A-B\gamma_5)u_i,
\end{eqnarray}
where $A$ and $B$ correspond to the parity-violating $S$-wave and parity-conserving $P$-wave amplitudes, respectively. They generally receive both factorizable and nonfactorizable contributions. In the 1990s various approaches such as the relativistic quark model, the pole model and current algebra were developed to describe the nonfactorizable effects in hadronic decays of charmed baryons \cite{Cheng:2021qpd}.

Besides the dynamical model calculations,  nonleptonic decays of charmed baryons also have been analyzed using the symmetry approaches such as the irreducible SU(3) approach (IRA) and the topological diagrammatic approach (TDA). These approaches, especially IRA, 
have became very popular in the past few years. 
Notice that early studies of the $SU(3)_F$ approach have overlooked the fact that charmed baryon decays are governed by several different partial-wave amplitudes which have distinct kinematic and dynamic effects.
After the improvement made in Ref. \cite{Geng:2019xbo}, now it became a common practice to perform a global fit of both $S$- and $P$-wave parameters to the data of branching fractions and decay asymmetries.  
Just like the case of hyperon decays, non-trival relative strong phases between $S$- and $P$-wave amplitudes may exist, but they are usually neglected in the model calculations.

There is one decay mode that deserves special attention, namely, the Cabibbo-favored decay $\Lambda_c^+\to\Xi^0 K^+$ which proceeds only through $W$-exchange. Early studies in 1990's indicated that its $S$- and $P$-wave amplitudes are very small due to strong cancellation between various terms (see e.g. Ref. \cite{Cheng:1993gf}). For example, the use of current algebra implies a vanishing $S$-wave in the SU(3) limit. Consequently, the calculated branching fraction is too small compared to experiment and the predicted $\alpha$ is zero owing to the vanishing $S$-wave amplitude. 
It is thus striking that the approach based on IRA tends to predict a large decay asymmetry close to unity \cite{Geng:2019xbo,Zhong:2022exp,Xing:2023dni}. It is also true in a revised pole model calculation in Ref. \cite{Zou:2019kzq}.
This long-standing puzzle was recently resolved by a new BESIII measurement  \cite{BESIII:2023wrw}. Not only the decay asymmetry $\alpha_{\Xi^0K^+}=0.01\pm0.16$ was found to be consistent with zero, but also the measured Lee-Yang parameter $\beta_{\Xi^0K^+}=-0.64\pm0.69$ was nonzero, implying a phase difference between $S$- and $P$-wave amplitudes, $\delta_P-\delta_S=-1.55\pm0.25$ or $1.59\pm0.25$ rad. Since $|\cos(\delta_P-\delta_S)|\sim 0.02$, this accounts for the smallness of $\alpha_{\Xi^0K^+}$. 
This phase-shift effect has been taken into account
in a recent IRA analysis of charmed baryon decays \cite{Geng:2023pkr}.

The TDA offers a systematic approach for studying the charm system. 
It has the advantage that it is more intuitive for both short-  and long-distance physics 
and easier to implement model calculations.
Since the TDA has been applied very successfully to charmed meson decays \cite{CC,Cheng:2016,Cheng:2024hdo}, 
it is conceivable that the same approach is applicable to the charmed baryon sector. 
For the recent analysis of hadronic charmed baryon decays in the TDA, see Refs. \cite{He:2018joe,Hsiao:2021nsc}.
In this work, we shall perform a study based on the TDA, incorporating
both partial-wave amplitudes and their phase shifts for the first time.

\vskip 0.2cm
\noindent{\it  2. Formalism---} 
\label{sec:Forma}
Since baryons are made of three quarks in contrast to two quarks for the mesons, the application of TDA to the baryon case will inevitably lead to some complications. For example, the symmetry of the quarks in flavor space could be different. The first analysis of two-body nonleptonic decays of antitriplet charmed baryons in TDA was carried out by Kohara \cite{Kohara:1991ug}. A subsequent study was given by Chau, Cheng and Tseng (CCT) in Ref. \cite{Chau:1995gk}. The difference between Kohara and CCT lies in the choice of the wave functions of octet baryons:
\begin{eqnarray} \label{eq:wf8}
|{\cal B}^{m,k}(8)\rangle&=&a|\chi^m(1/2)_{A_{12}}\rangle|\psi^k(8)_{A_{12}}\rangle
\nonumber \\ &+&  b|\chi^m(1/2)_{S_{12}}\rangle|\psi^k(8)_{S_{12}}\rangle  
\end{eqnarray}
with $|a|^2+|b|^2=1$ in Ref. \cite{Chau:1995gk}, and
\begin{eqnarray}
|\tilde{\cal B}^{m,k}(8)\rangle=\alpha|\chi^m(1/2)_{A_{12}}\rangle|\psi^k(8)_{A_{12}}\rangle
\nonumber \\+ \beta|\chi^m(1/2)_{A_{23}}\rangle|\psi^k(8)_{A_{23}}\rangle
\end{eqnarray}
in Ref. \cite{Kohara:1991ug}, where $\chi^m(1/2)_{A,S}$ are the spin parts of the wave function defined in Eq. (23) of Ref. \cite{Chau:1995gk} and
\begin{eqnarray}
|\psi^k(8)_{A_{12}}\rangle &=& \sum_{q_a,q_b,q_c}|[q_aq_b]q_c\rangle \langle[q_aq_b]q_c
|\psi^k(8)_{A_{12}}\rangle, \nonumber \\
|\psi^k(8)_{S_{12}}\rangle &=& \sum_{q_a,q_b,q_c}|\{q_aq_b\}q_c\rangle \langle\{q_aq_b\}q_c
|\psi^k(8)_{S_{12}}\rangle, 
\end{eqnarray}
denote the octet baryon states that are antisymmetric and symmetric in the first two quarks, respectively. As shown explicitly in Ref. \cite{Kohara:1997nu}, physics is independent of the convention one chooses. The TDA amplitudes expressed in the schemes with ${\cal B}^{m,k}(8)$ and $\tilde {\cal B}^{m,k}(8)$ are equivalent. Nevertheless, we prefer to use the bases $\psi^k(8)_{A_{12}}$ and $\psi^k(8)_{S_{12}}$ as they are orthogonal to each other, while $\psi^k(8)_{A_{12}}$ and $\psi^k(8)_{A_{23}}$ are not. 

\begin{figure}[t]
	\includegraphics[scale=0.60]{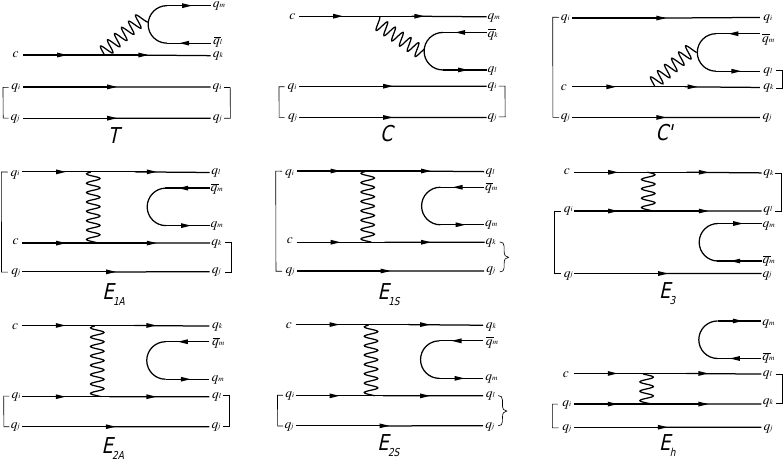}
	\caption{Topological diagrams contributing to ${\cal B}_c(\bar 3)\to {\cal B}(8)M(8+1)$ decays.}
	\label{Fig:TopDiag}
\end{figure}

In terms of the octet baryon wave functions given in Eq. (\ref{eq:wf8}), the relevant topological diagrams for the decay ${\cal B}_c(\bar 3)\to {\cal B}(8)M(8+1)$ are depicted in Fig. \ref{Fig:TopDiag}: the external $W$-emission, $T$; the internal $W$-emission $C$; the inner $W$-emission $C'$; $W$-exchange diagrams $E_{1A}$, $E_{1S}$, $E_{2A}$, $E_{2S}$, 
$E_3$ and the hairpin diagram $E_h$. 
The decay amplitudes of ${\cal B}_c(\bar 3)\to {\cal B}(8)M(8+1)$ in TDA have the expressions:  
\begin{equation}
\label{Eq:TDAamp}
\begin{aligned}
 \mathcal{A}_{\rm T D A}
=
& \quad T ({\mathcal{B}}_c)^{i j} H_l^{k m}\left(\mathcal{B}_8\right)_{i j k} M_m^l \\
&+C (\mathcal{B}_c)^{i j} H_k^{m l}\left(\mathcal{B}_8\right)_{i j l} M_m^k\\
 &+ C' (\mathcal{B}_c)^{i j} H_m^{k l}\left(\mathcal{B}_8\right)_{klj} M_i^m\\
& +E_{1A} (\mathcal{B}_c)^{i j} H_i^{k l}\left(\mathcal{B}_8\right)_{jkm} M_l^m\\
&+ E_{1S} (\mathcal{B}_c)^{i j} H_i^{k l}M_l^m \left[\left(\mathcal{B}_8\right)_{jmk} 
+\left(\mathcal{B}_8\right)_{kmj} \right]\\
&+E_{2A} (\mathcal{B}_c)^{i j} H_i^{k l}\left(\mathcal{B}_8\right)_{jlm} M_k^m
\\
&+ E_{2S} (\mathcal{B}_c)^{i j} H_i^{k l} M_k^m\left[\left(\mathcal{B}_8\right)_{jml}
+ \left(\mathcal{B}_8\right)_{lmj} \right]\\
&   +E_{3} (\mathcal{B}_c)^{i j} H_i^{k l}\left(\mathcal{B}_8\right)_{klm} M_j^m\\
&   +E_{h} (\mathcal{B}_c)^{i j} H_i^{k l}\left(\mathcal{B}_8\right)_{klj} M_m^m, 
\end{aligned}
\end{equation}
where $(\mathcal{B}_c)^{ij}$ is an antisymmetric baryon matrix standing for antitriplet charmed baryons, $(\mathcal{B}_8)^i_j$ and  $M^{i}_{j}$ represent octet baryons and nonet mesons, respectively, see e.g. Refs. \cite{He:2018joe,Fanrong} for their explicit expressions. 
Based on the convention of $(\mathcal{B}_8)^i_j$, we adopt the notation 
$(\mathcal{B}_8)_{i j k} = \epsilon_{ijl} (\mathcal{B}_8)^{l}_{k}$ to describe the baryon octet.
The $H$ matrix is related to the CKM mixing matrix
with the non-vanishing elements:
$H_{2}^{31}=V_{cs}^*V_{ud}$, $H_{3}^{31}=V_{cs}^*V_{us}$, $H_{2}^{21}=V^*_{cd}V_{ud}$ and $H_{3}^{21}=V^*_{cd}V_{us}$.

Note that there are 7 topological diagrams in Ref. \cite{He:2018joe} 
corresponding to 19 TDA amplitudes. However, in diagrams $T$ and $C$, the two spectator quarks $q_i$ and $q_j$ are antisymmetric in flavor. Moreover, the final-state quarks $q_l$ and $q_k$  in topological diagrams $C'$, $E_3$ and $E_h$ must be antisymmetric in flavor owing to the K\"orner-Pati-Woo (KPW) theorem which states that the quark pair in a baryon produced by weak interactions is required to be antisymmetric in flavor \cite{Korner:1970xq}. Consequently, the number of TDA amplitudes is reduced from 19 to 11.  Next, one needs to consider the symmetric and antisymmetric parts of $E_{1,2}$ diagrams separately. Moreover, the final-state quarks $q_l$ and $q_k$ in topological diagrams $E_{1A,1S}$ and $E_{2A,2S}$
are required by the KPW theorem be antisymmetric in flavor.
We thus led to $E_{2A}=-E_{1A}$ and $E_{2S}=-E_{1S}$.
As a result,  the number of independent topological diagrams in Fig. \ref{Fig:TopDiag} and TDA amplitudes is 7. 

Working out Eq. (\ref{Eq:TDAamp}) for ${\cal B}_c(\bar 3)\to {\cal B}(8)M(8+1)$ decays,  we obtain the TDA decay amplitudes listed in Table \ref{tab:CFamp}. 
For complete TDA amplitudes of singly- and doubly-Cabibbo-suppressed decays, see Ref. \cite{Fanrong}.
The expressions of TDA amplitudes agree with CCT \cite{Chau:1995gk} through the following relations:
\begin{eqnarray}
&& {\cal A}_A=-4T, \quad {\cal B}_A=2C', \quad {\cal B}'_A=-4C,   \nonumber \\
&& {\cal C}_{1A}=2E_3,  \quad {\cal C}_{2A}=-{\cal C}'_{A}=2E_{1A}, \\
&& {\cal C}_{2S}=-{\cal C}'_S=-2\sqrt{3}E_{1S},  \nonumber
\end{eqnarray}
where ${\cal A}_A$, ${\cal B}_A$, ${\cal B}'_A$, ${\cal C}_{1A}$, ${\cal C}_{2A}$, ${\cal C}'_{A}$, 
${\cal C}_{2S}$ and ${\cal C}'_S$ are the TDA amplitudes defined in Ref. \cite{Chau:1995gk}. The agreement is non-trivial in view of the different approaches adopted in Ref. \cite{Chau:1995gk} and here. 
Among the 7 TDA amplitudes given in Eq. (\ref{Eq:TDAamp}), there still exist 2 redundant degrees of freedom through the redefinition \cite{Chau:1995gk}:
\begin{eqnarray}
\label{eq:tildeTDA}
&& \tilde T=T-E_{1S}, \quad \tilde C=C+E_{1S},\quad \tilde C'=C'-2E_{1S},   \nonumber \\
&&  \tilde E_h=E_h+2E_{1S} \quad \tilde E_1=-\tilde E_2=E_{1A}+E_{1S}-E_3.   
\end{eqnarray}
It is clear that the minimum set of the topological amplitudes in TDA is 5. This is in agreement with the number of tensor invariants found in IRA \cite{Geng:2023pkr}.

Many relations can be easily derived from Table \ref{tab:CFamp}, which will be discussed in detail in Ref. \cite{Fanrong}. One of them is 
\begin{eqnarray}
\label{eq:sumrule}
{\tau_{\Lambda_c^+}\over\tau_{\Xi_c^0}}\B(\Xi_c^0\to\Xi^-\pi^+) &=&\B(\Lambda_c^+\to \Sigma^0\pi^+)+3\B(\Lambda_c^+\to \Lambda\pi^+) \nonumber \\
&-&{1\over \sin^2\theta_C}\B(
\Lambda_c^+\to n\pi^+).
\end{eqnarray}
This relation in agreement with Ref. \cite{Geng:2023pkr}  is very useful in constraining the branching fraction of $\Xi_c^0\to\Xi^-\pi^+$.

\begin{table}[t]
\centering
\caption{TDA amplitudes for CF (upper part) and several SCS (lower part) ${\cal B}_c(\bar 3)\to {\cal B}(8)M(8+1)$ decays.  The amplitudes for CF (SCS) decays have to be multiplied by the matrix element $H^{31}_2$ ($H_3^{31}$ or $H_2^{21}$, depending on the process). For the $\eta$ and $\eta'$ mesons, we take $\eta=\cos\theta \eta_8-\sin\theta\eta_1, \eta'=\sin\theta\eta_8+\cos\theta\eta_1$ with $\theta\approx -15^\circ$.}
\label{tab:CFamp}
\begin{tabular}{ll}
\hline
Channel & ~~~~~~~~TDA   \\
\hline
$\Lambda_c^+ \to \Lambda \pi^+$ &  
$\frac{1}{\sqrt{6}}(-4T+C'+E_{1A}+3E_{1S}-E_{3})$ \\

$\Lambda_c^+ \to \Sigma^0 \pi^+$ &  
$\frac{1}{\sqrt{2}}(-C'-E_{1A}+E_{1S}+E_{3})$\\

$\Lambda_c^+ \to \Sigma^+ \pi^0$ &  
$\frac{1}{\sqrt{2}}(C'+E_{1A}-E_{1S}-E_{3})$\\

$\Lambda_c^+ \to \Sigma^+ \eta_{8}$ &  
$\frac{1}{\sqrt{6}}(-C'+E_{1A}+3E_{1S}-E_{3})$ \\

$\Lambda_c^+ \to \Sigma^+ \eta_{1}$ &  
$\frac{1}{\sqrt{3}}(-C'+E_{1A}-3E_{1S}-E_{3}-3E_h)$\\

$\Lambda_c^+ \to \Xi^0 K^+$&
$E_{1A}+E_{1S}-E_{3}$\\

$\Lambda_c^+ \to p \bar{K}^0$&
$2C+2E_{1S}$\\

$\Xi_c^0 \to \Lambda \bar{K}^0$&
$\frac{1}{\sqrt{6}}(2C-C'-E_{1A}+3E_{1S}+E_{3})$ \\

$\Xi_c^0 \to \Sigma^0 \bar{K}^0$&
$\frac{1}{\sqrt{2}}(2C+C'+E_{1A}+E_{1S}-E_{3})$
\\

$\Xi_c^0 \to \Sigma^+ K^-$&
$-E_{1A}-E_{1S}+E_{3}$\\

$\Xi_c^0 \to \Xi^0 \pi^0$&
$\frac{1}{\sqrt{2}}(-C'+2E_{1S})$\\



$\Xi_c^0 \to \Xi^0 \eta_8$&
$\frac{1}{\sqrt{6}}(C'+2E_{1A}-2E_{3})$\\

$\Xi_c^0 \to \Xi^0 \eta_1$&
$\frac{1}{\sqrt{3}}(C'+3E_{1S}-E_{1A}+E_{3}+ 3E_h)$\\

$\Xi_c^0 \to \Xi^- \pi^+$&
$2T-2E_{1S}$\\

$\Xi_c^+ \to \Sigma^+ \bar{K}^0$&
$-2C-C'$\\

$\Xi_c^+ \to \Xi^0 \pi^+$&
$-2T+C'$\\
\hline

$\Lambda_c^+ \to \Lambda K^+$ &  
$\frac{1}{\sqrt{6}}(-4T+C'-2E_{1A}+2E_{3})$\\

$\Lambda_c^+ \to \Sigma^0 K^+$ & 
$\frac{1}{\sqrt{2}}(-C'+2E_{1S})$\\

$\Lambda_c^+ \to \Sigma^+ K^0$ &
$-C'+2E_{1S}$\\

$\Lambda_c^+ \to p \pi^0$ &
$\frac{1}{\sqrt{2}}(2C+C'+E_{1A}+E_{1S}-E_{3})$\\

$\Lambda_c^+ \to p \eta_8$ &  
$\frac{1}{\sqrt{6}}(-6C+C'+E_{1A}-3E_{1S}-E_{3})$\\

$\Lambda_c^+ \to p \eta_1$&
$\frac{1}{\sqrt{3}}(-C'+E_{1A}-3E_{1S}-E_{3}-3E_h)$\\

$\Lambda_c^+ \to n \pi^+$&
$-2T+C'+E_{1A}+E_{1S}-E_{3}$ \\

$\Xi_c^0 \to \Xi^- K^+$&
$2T-2E_{1S}$\\

\hline
\end{tabular}
\end{table}

\vskip 0.2cm
\noindent{\it  3. Numerical Analysis and Results---} 
\label{sec:Num}
As there are 5 independent tilde TDA amplitudes given in Eq. (\ref{eq:tildeTDA}),  we have totally 20 unknown parameters to describe the magnitudes and the phases of the respective $S$- and $P$-wave amplitudes, namely, $|\tilde T|_Se^{i\delta_S^{\tilde T}}$,       
$|\tilde C|_Se^{i\delta_S^{\tilde C}}$, $\cdots$, $|\tilde E_h|_Se^{i\delta_S^{\tilde E_h}}$, $|\tilde T|_P e^{i\delta_P^{\tilde T}}$,
$\cdots$, $|\tilde E_h|_P e^{i\delta_P^{\tilde E_h}}$, collectively denoted by $|X_i|_Se^{i\delta^{X_i}_S}$ and $|X_i|_Pe^{i\delta^{X_i}_P}$,
where the subscripts $S$ and $P$ denote the $S$- and $P$-wave components of each TDA amplitude. 
Since there is an overall phase which can be omitted, we shall set $\delta_S^{\tilde T}=0$. Hence,  we are left with 19 parameters. 
Notice that the number of available experimental observables has increased to 30 by the end of 2023.
To pursue a set of proper parameters, 
the $\chi^2$ function in the following maximum likelihood analysis is defined as
\begin{equation}
\chi^2=\left[\mathcal{O}_{\text{theor}}(c_i)-\mathcal{O}_{\text{expt}}\right]^{\text{T}}\Sigma^{-1}
\left[\mathcal{O}_{\text{theor}}(c_i)-\mathcal{O}_{\text{expt}}\right],
\end{equation}
in which $c_i$ are the fitted $19$ input parameters, $\mathcal{O}_{\text{theor}, \text{expt}}$
stand for the $30$ theoretical and experimental observables.
The  $30$-dimensional general error matrix $\Sigma$ can be taken diagonal by
neglecting correlations among different observables and only 
incorporating pure experimental errors here.
In addition to the latest PDG values \cite{Workman:2022ynf} adopted as 
partial inputs in $\chi^2$,  
more $\Lambda_c^+$
related data have been supplied by BESIII in 2023, including branching fractions of 
$\Lambda_c^+\to p \pi^0$ \cite{BESIII:2023uvs} and 
$\Lambda_c^+ \to p \eta$ \cite{BESIII:2023wrw}.
Likewise, Belle has also contributed the recent measurements to $\Xi_c^{0}$, 
such as $\Xi_c^0\to \Xi^- \pi^+$ \cite{Belle:2018kzz} and $\Xi_c^0\to \Lambda^0 K_S, \Sigma^0 K_S, \Sigma^+ K^-$ \cite{Belle:2021avh}.

\begin{table}[t]
\caption{ Fitted tilde TDA amplitudes collectively denoted by $X_i$. 
} 
\label{tab:coeff}
\vspace{-0.1cm}
\begin{center}
\renewcommand\arraystretch{1}
\begin{tabular}
{c| c r r r }
\hline
&$|X_i|_S$&$|X_i|_P$~~~ &$\delta^{X_i}_{S}$~~~~~ &$\delta^{X_i}_{P}$~~~~ \\
&\multicolumn{2}{c}{$(10^{-2}G_{F}~{\rm GeV}^2)$}&\multicolumn{2}{c}{$(\text{in radian})$} \\
\hline
$\Tilde{T}$&
$2.37\pm0.41$&$16.56\pm0.69$&
-- ~~~~~~~&$2.76\pm0.32$\\
$\Tilde{C}$&
$1.04\pm1.08$&$13.82\pm0.58$&
$-1.97\pm0.79$&$-0.37\pm0.44$\\
$\Tilde{C'}$&
$2.59\pm0.95$&$24.97\pm1.67$&
$0.29\pm0.19$&$2.86\pm0.36$\\
$\Tilde{E_{1}}$&
$4.10\pm0.20$&$2.56\pm2.21$&
$1.18\pm0.38$&$-0.96\pm0.43$\\
$\Tilde{E_{h}}$&
$1.54\pm1.22$&$19.16\pm3.00$&
$-1.35\pm0.60$&$0.37\pm0.41$\\
\hline
\end{tabular}
\end{center}
\end{table}

In terms of the $S$- and $P$-wave amplitudes given in Eq. (\ref{eq:A&B}) 
and their phases $\delta_S$ and $\delta_P$, respectively, the decay rate and decay asymmetries read
\begin{eqnarray}
&&\Gamma = \frac{p_c}{8\pi}\frac{(m_i+m_f)^2-m_P^2}{m_i^2}\left(|A|^2
+ \kappa^2|B|^2\right),\nonumber \\
&& \alpha=\frac{2\kappa |A^*B|\cos(\delta_P-\delta_S)}{|A|^2+\kappa^2 |B|^2},~~
\beta=\frac{2\kappa |A^*B|\sin(\delta_P-\delta_S)}{|A|^2+\kappa^2 |B|^2}, \nonumber \\
&& \gamma=\frac{|A|^2-\kappa^2 |B|^2}{|A|^2+\kappa^2 |B|^2}, 
\label{eq:kin}
\end{eqnarray}
where  $p_c$ is the c.m. three-momentum in the rest
frame of the initial baryon and the auxiliary parameter $\kappa$ is 
defined as $\kappa=p_c/(E_f+m_f)$. The available experimental data are collected in Table \ref{tab:fitCF} below. Note that for $\Xi_c^{0}$ decays, many of the modes are measured relative to $\Xi_c^0\to \Xi^-\pi^+$; that is, ${\cal R}_X\equiv \B(\Xi_c^0\to X)/\B(\Xi_c^0\to \Xi^-\pi^+)$ for $X=\Xi^-K^+, \Lambda^0K_S^0, \Sigma^0K_S^0$ and $\Sigma^+K^-$.

\begin{table*}[t]
\caption{
The fit results based on the tilde TDA. $S$- and $P$-wave amplitudes are in units of $10^{-2}G_F~{\rm GeV}^2$ and $\delta_P-\delta_S$ in radian. 
}
\label{tab:fitCF}
\centering
\resizebox{\textwidth}{!} 
{
\begin{tabular}
{ l |c r r r r|c c}
\hline
\hline
Channel
&$10^{2}\mathcal{B}$&$\alpha$~~~~~~ &$|A|$~~~~~ &$|B|$~~~~~ &$\delta_P-\delta_S$~~~ &
$10^{2}\mathcal{B}_\text{exp}$
&$\alpha_\text{exp}$\\
\hline
$\Lambda_c^+\to\Lambda^0\pi^+$&$1.31\pm0.05$&$-0.76\pm0.01$&$2.76\pm0.25$&$16.96\pm0.39$&$-2.92\pm0.29$&$1.29\pm{0.05}$&$-0.76\pm0.01$\cite{Workman:2022ynf,Belle:2022uod}\\
$\Lambda_c^+\to\Sigma^0\pi^+$&$1.26\pm0.05$&$-0.48\pm0.02$&$4.07\pm0.86$&$15.48\pm2.29$&$2.08\pm0.04$&$1.27\pm{0.06}$&$-0.47\pm0.03$\cite{Workman:2022ynf,Belle:2022uod}\\
$\Lambda_c^+\to\Sigma^+ \pi^0$&$1.27\pm0.05$&$-0.48\pm0.02$&$4.07\pm0.86$&$15.48\pm2.29$&$2.08\pm1.15$&$1.25\pm0.09$&$-0.49\pm0.03$\cite{Workman:2022ynf,Belle:2022bsi}\\
$\Lambda_c^+\to\Sigma^+ \eta$&$0.33\pm0.04$&$-0.93\pm0.05$&$2.30\pm0.35$&$9.48\pm1.16$&$-2.80\pm0.16$&$0.32\pm0.04$\cite{Workman:2022ynf, Belle:2022bsi}&$-0.99\pm0.06$\cite{Belle:2022bsi}\\
$\Lambda_c^+\to\Sigma^+ \eta'$&$0.39\pm0.11$&$-0.45\pm0.07$&$3.81\pm1.44$&$23.04\pm3.84$&$-4.25\pm0.08$&$0.44\pm0.15$\cite{Workman:2022ynf, Belle:2022bsi}&$-0.46\pm0.07$\cite{Belle:2022bsi}\\
$\Lambda_c^+\to\Xi^0 K^+$&$0.41\pm0.03$&$-0.16\pm0.13$&$3.89\pm0.19$&$2.43\pm2.10$&$-2.15\pm0.65$&$0.55\pm0.07$&$0.01\pm0.16$\cite{BESIII:2023wrw}\\
$\Lambda_c^+\to\Lambda^0 K^+$&$0.0639\pm0.0030$&$-0.56\pm0.05$&$1.09\pm0.18$&$3.32\pm0.59$&$2.17\pm0.06$&$0.0635\pm0.0031$\cite{Workman:2022ynf,Belle:2022uod}&$-0.585\pm0.052$\cite{Belle:2022uod}\\
$\Lambda_c^+\to\Sigma^0 K^+$&$0.0376\pm0.0032$&$-0.54\pm0.08$&$0.40\pm0.15$&$3.86\pm0.26$&$2.56\pm0.44$&$0.0382\pm0.0051$\cite{Workman:2022ynf,Belle:2022uod}&$-0.55\pm0.20$\cite{Belle:2022uod}\\
$\Lambda_c^+\to\Sigma^+K_S$&$0.0377\pm0.0032$&$-0.54\pm0.08$&$0.40\pm0.15$&$3.86\pm0.26$&$2.56\pm0.44$&$0.047\pm0.014$&\\
$\Lambda_c^+\to n\pi^+$&$0.063\pm0.009$&$-0.78\pm0.13$&$1.01\pm0.14$&$2.43\pm0.38$&$3.81\pm0.30$&$0.066\pm0.013$&\\
$\Lambda_c^+\to p\pi^0$&$0.0176\pm0.0034$&$-0.11\pm0.75$&$0.64\pm0.13$&$0.94\pm0.66$&$4.59\pm1.70$&$0.0156_{-0.0061}^{+0.0075}$\cite{BESIII:2023uvs}&\\
$\Lambda_c^+\to p K_S$&$1.57\pm0.07$&$0.01\pm0.31$&$1.41\pm1.51$&$18.68\pm0.79$&$1.54\pm0.82$&$1.59\pm0.07$&$0.18\pm0.45$\\
$\Lambda_c^+\to p\eta$&$0.151\pm0.008$&$0.07\pm0.37$&$1.01\pm0.53$&$5.46\pm0.67$&$1.48\pm0.45$&$0.149\pm0.008$\cite{Workman:2022ynf,BESIII:2023uvs,BESIII:2023ooh}&\\
$\Lambda_c^+\to p\eta'$&$0.052\pm0.009$&$-0.54\pm0.19$&$0.77\pm0.30$&$4.72\pm0.73$&$2.29\pm0.13$&$0.049\pm0.009$&\\
$\Xi_c^0\to\Xi^- \pi^+$&$2.83\pm0.10$&$-0.72\pm0.03$&$4.51\pm0.79$&$31.47\pm1.31$&$2.76\pm0.32$&$1.80\pm0.52$\cite{Belle:2018kzz}&$-0.64\pm0.05$\\
$\Xi_c^+\to\Xi^0\pi^+$&$0.9\pm0.2$&$-0.93\pm0.07$&$2.27\pm0.30$&$8.21\pm1.16$&$-3.5\pm0.23$&$1.6\pm0.8$&\\
\hline
Channel
&$10^{2}\mathcal{R}_X$&$\alpha$&$|A|$&$|B|$~~~~~ &$\delta_P-\delta_S$~~~ &
$10^{2}(\mathcal{R}_X)_\text{exp}$
&$\alpha_\text{exp}$\\
\hline
$\Xi_c^0\to\Xi^-K^+$&$4.10\pm0.05$&$-0.76\pm0.03$&$1.04\pm0.18$&$7.25\pm0.30$&$2.76\pm0.32$&$2.75\pm0.57$&\\
$\Xi_c^0\to\Lambda K_S^0$&$24.0\pm1.0$&$-0.23\pm0.19$&$2.06\pm0.87$&$13.59\pm1.13$&$1.89\pm0.32$&$22.9\pm1.4$\cite{Belle:2021avh}&\\
$\Xi_c^0\to\Sigma^0 K_S^0$&$3.9\pm0.7$&$0.01\pm0.65$&$1.92\pm0.43$&$3.47\pm2.00$&$4.72\pm1.67$&$3.8\pm0.7$\cite{Belle:2021avh}&\\
$\Xi_c^0\to\Sigma^+K^-$&$13.0\pm1.1$&$-0.21\pm0.17$&$3.89\pm0.19$&$2.43\pm2.10$&$4.13\pm0.65$&$12.3\pm1.2$\cite{Belle:2021avh}&\\
\hline
\hline
\end{tabular}
}
\end{table*}

In practice, 
we shall make use of the package \texttt{iminuit} \cite{iminuit, James:1975dr}
to search for $\chi^2_{\text{min}}$ together with
its corresponding fitted parameters $c_i$ exhibited in 
Table \ref{tab:coeff}, and generate the covariance matrix 
among  parameters which further helps 
predict physical observables. 
The fit branching fractions, decay asymmetries, the magnitudes of $S$ and $P$ waves and their phase shifts are shown in Tables  \ref{tab:fitCF} and \ref{tab:fitother}. Owing to the space limit,  the fitted parameters $\beta$ and $\gamma$ will be presented in Ref. \cite{Fanrong}.
We see that the fitting  results for the branching fractions and decay asymmetries are in good agreement with experiment except for the following three modes: $\Xi_c^0\to \Xi^-\pi^+$, $\Lambda_c^+\to \Xi^0 K^+$ and the ratio ${\cal R}_{\Xi^-K^+}$. The $\chi^2$ value is 2.0 per degree of freedom. 
Our predicted 
branching fraction of $(2.83\pm0.10)\%$ for
$\Xi_c^0\to \Xi^-\pi^+$ is noticeably higher than the value of $(1.80\pm0.52)\%$ measured by Belle \cite{Belle:2018kzz}. A similar result of $(2.72\pm0.09)\%$ has also been obtained in Ref. \cite{Geng:2023pkr}. 
Indeed, the sum-rule relation derived in Eq. (\ref{eq:sumrule}) is well satisfied by the fitting branching fractions listed in Table \ref{tab:fitCF}.

As for the ratio ${\cal R}_{\Xi^-K^+}$, we see from Table \ref{tab:CFamp} that in the
SU(3) limit, one will have ${\cal R}_{\Xi^-K^+}=\sin^2\theta_C$ which is equal to 0.045 after taking into account the phase-space difference between  $\Xi_c^0\to \Xi^-K^+$ and $\Xi_c^0\to \Xi^-\pi^+$. The current measurement is $0.0275\pm0.0057$ which is away from the SU(3) expectation by $2\sigma$. Since both modes proceed through the topological diagrams $T$ and $E_{1S}$ with the combination $2T-2E_{1S}=2\tilde T$, it is conceivable that SU(3) breaking in the external $W$-emission $T$ and especially in $W$-exchange $E_{1S}$ will account for the discrepancy between theory and experiment.

If the pseudoscalar meson in the final-state is an SU(3) flavor-singlet $\eta_1$, it will receive an additional contribution from the hairpin diagram $E_h$ in Fig. \ref{Fig:TopDiag}. Numerically, we find that the combination $\tilde E_h=E_h+2E_{1S}$
which contributes only to $\eta_1$ is sizable (see Table \ref{tab:coeff}). Recall that in the charmed meson sector there is a strong indication of the hairpin effect in the decay $D_s^+\to \rho^+\eta'$ (see e.g. Ref. \cite{Cheng:2024hdo}).

For the decay $\Lambda_c^+\to \Xi^0 K^+$, we find
$|A|=3.89\pm0.19$,  $|B|=2.43\pm2.10$, both
in units of $10^{-2} G_F~{\rm GeV}^2$, and the phase difference
$\delta_P-\delta_S=-2.15\pm 0.65$ rad. Our fit $\B(\Lambda_c^+\to \Xi^0 K^+)=(0.41\pm0.03)\%$ is slightly smaller than the measured value of $(0.55\pm0.07)\%$. Nevertheless, the fit $\alpha_{\Xi^0K^+}=-0.16\pm0.13$ is consistent with the measurement of  $0.01\pm0.16$ \cite{BESIII:2023wrw}. 
Our results are very close to those obtained in Ref. \cite{Geng:2023pkr}:
$\B(\Lambda_c^+\to \Xi^0 K^+)=(0.40\pm0.03)\%$, $\alpha_{\Xi^0K^+}=-0.15\pm0.14$ and $\delta_P-\delta_S=-2.06\pm 0.50$ rad.

We have checked that if we set $\delta_{S}=\delta_P=0$ from the outset and keep the measured $\alpha_{\Xi^0K^+}$ as an input, the fit $\B(\Lambda_c^+\to \Xi^0 K^+)$ of order $1\times 10^{-3}$ will be too small compared to experiment. On the contrary, if the input of $(\alpha_{\Xi^0K^+})_{\rm exp}$ is removed, the fit $\alpha_{\Xi^0K^+}$ will be of order 0.95\,. Hence, we conclude that it is inevitable to introduce the phase shifts to accommodate the data.

As noticed in passing, the minimum set of the topological amplitudes in the TDA and the number of tensor invariants in the IRA are the same. 
The  relations between 
the TDA and IRA will be discussed in more detail in Ref. \cite{Fanrong}.

\begin{table*}[t]
\caption{
Same as Table \ref{tab:fitCF} except for yet-measured  modes.
}
\label{tab:fitother}
\resizebox{\textwidth}{!} 
{
\centering
\begin{tabular}
{ l | r r c c r
| l|r r c c r}
\hline
\hline
Channel&
$10^{3}\mathcal{B}$~~~~ &$\alpha$~~~~~~
&$|A|$ & $|B|$ & $\delta_P-\delta_S$~~~
& Channel& $10^{4}\mathcal{B}$~~~~ &$\alpha$~~~~~~
&$|A|$ & $|B|$ &$\delta_P-\delta_S$~~~
\\
\hline
$\Lambda_c^{+} \rightarrow p K_L$&$15.54\pm0.65$&$-0.03\pm0.28$&$1.38\pm1.40$&$18.48\pm0.78$&$1.65\pm0.80$&$\Lambda_c^{+} \rightarrow n K^{+}$&$0.12\pm0.03$&$-0.88\pm0.45$&$0.12\pm0.02$&$0.44\pm0.06$&$-3.50\pm0.23$\\
\hline
$\Xi_c^{+} \rightarrow \Sigma^{+} K_S$&$1.64\pm2.54$&$0.46\pm0.43$&$1.36\pm1.01$&$1.69\pm4.46$&$-0.42\pm6.12$&$\Xi_c^{+} \rightarrow \Lambda^0 \pi^{+}$&$3.55\pm1.15$&$0.21\pm0.28$&$0.24\pm0.12$&$1.70\pm0.28$&$-1.25\pm0.48$\\
$\Xi_c^{+} \rightarrow p K_{S / L}$&$2.00\pm0.20$&$-0.38\pm0.07$&$0.40\pm0.15$&$3.86\pm0.26$&$2.56\pm0.44$&$\Xi_c^{+} \rightarrow n \pi^{+}$&$0.34\pm0.04$&$-0.28\pm0.22$&$0.21\pm0.01$&$0.13\pm0.11$&$-2.15\pm0.65$\\
$\Xi_c^{+} \rightarrow \Sigma^{+} \pi^0$&$2.16\pm0.23$&$-0.07\pm0.37$&$0.96\pm0.32$&$3.96\pm0.52$&$1.64\pm0.40$&$\Xi_c^{+} \rightarrow \Sigma^0 K^{+}$&$1.22\pm0.05$&$-0.68\pm0.03$&$0.17\pm0.03$&$1.18\pm0.05$&$-3.53\pm0.32$\\
$\Xi_c^{+} \rightarrow \Sigma^{+} \eta$&$0.49\pm0.34$&$-0.41\pm0.70$&$0.52\pm0.20$&$2.10\pm1.38$&$-2.01\pm0.78$&$\Xi_c^{+} \rightarrow p \pi^0$&$0.17\pm0.02$&$-0.28\pm0.22$&$0.15\pm0.01$&$0.09\pm0.08$&$-2.15\pm0.65$\\
$\Xi_c^{+} \rightarrow \Sigma^{+} \eta'$&$3.32\pm0.62$&$-0.30\pm0.12$&$1.15\pm0.61$&$10.11\pm1.38$&$-4.32\pm0.21$&$\Xi_c^{+} \rightarrow p \eta$&$2.22\pm0.37$&$-0.38\pm0.07$&$0.20\pm0.06$&$1.14\pm0.15$&$2.17\pm5.08$\\
$\Xi_c^{+} \rightarrow \Sigma^0 \pi^{+}$&$3.12\pm0.13$&$-0.59\pm0.04$&$1.13\pm0.24$&$4.80\pm0.56$&$2.26\pm0.08$&$\Xi_c^{+} \rightarrow p \eta'$&$1.99\pm0.42$&$-0.40\pm0.17$&$0.20\pm0.07$&$1.06\pm0.17$&$-4.13\pm0.12$\\
$\Xi_c^{+} \rightarrow \Xi^0 K^{+}$&$1.00\pm0.16$&$-0.73\pm0.12$&$1.01\pm0.14$&$2.43\pm0.38$&$-2.47\pm0.21$&$\Xi_c^{+} \rightarrow \Lambda^0 K^{+}$&$0.35\pm0.05$&$-0.41\pm0.13$&$0.20\pm0.02$&$0.30\pm0.12$&$2.09\pm0.20$\\
\hline
$\Xi_c^0 \rightarrow \Sigma^0 K_L$&$1.12\pm0.18$&$-0.20\pm0.66$&$2.02\pm0.37$&$2.35\pm2.24$&$-1.88\pm1.06$&$\Xi_c^0 \rightarrow p K^{-}$&$1.99\pm0.21$&$-0.27\pm0.22$&$0.90\pm0.04$&$0.56\pm0.48$&$4.13\pm0.65$\\
$\Xi_c^0 \rightarrow \Xi^0 \pi^0$&$7.45\pm0.64$&$-0.51\pm0.08$&$1.74\pm0.64$&$16.78\pm1.12$&$2.56\pm0.44$&$\Xi_c^0 \rightarrow n K_{S / L}$&$7.41\pm0.79$&$-0.43\pm0.05$&$0.94\pm0.20$&$3.56\pm0.53$&$2.08\pm0.04$\\
$\Xi_c^0 \rightarrow \Xi^0 \eta$&$2.87\pm0.69$&$0.08\pm0.20$&$3.12\pm0.46$&$6.61\pm2.15$&$1.48\pm0.23$&$\Xi_c^0 \rightarrow \Lambda^0 \pi^0 $&$1.12\pm0.36$&$-0.61\pm0.21$&$0.31\pm0.13$&$1.54\pm0.22$&$3.90\pm5.70$\\
$\Xi_c^0 \rightarrow \Xi^0 \eta'$&$13.98\pm2.17$&$-0.58\pm0.08$&$4.87\pm1.36$&$23.13\pm3.89$&$2.22\pm0.08$&$\Xi_c^0 \rightarrow n \pi^0 $&$0.06\pm0.01$&$-0.28\pm0.22$&$0.15\pm0.01$&$0.09\pm0.08$&$4.13\pm0.65$\\
$\Xi_c^0 \rightarrow \Lambda^0 K_L$&$6.80\pm0.22$&$-0.27\pm0.17$&$2.36\pm0.88$&$14.13\pm1.29$&$1.92\pm0.27$&$\Xi_c^0 \rightarrow \Lambda^0 \eta$&$4.56\pm0.97$&$0.22\pm0.23$&$0.19\pm0.23$&$4.00\pm0.48$&$0.57\pm4.95$\\
$\Xi_c^0 \rightarrow \Sigma^{+} \pi^{-}$&$0.22\pm0.02$&$-0.23\pm0.18$&$0.90\pm0.04$&$0.56\pm0.48$&$-2.15\pm0.65$&$\Xi_c^0 \rightarrow \Lambda^0 \eta'$&$6.85\pm0.99$&$-0.64\pm0.11$&$1.21\pm0.35$&$5.94\pm0.93$&$-3.98\pm0.09$\\
$\Xi_c^0 \rightarrow \Sigma^0 \pi^0 $&$0.34\pm0.04$&$-0.02\pm0.28$&$0.22\pm0.25$&$3.26\pm0.22$&$-1.62\pm0.74$&$\Xi_c^0 \rightarrow \Sigma^{-} K^{+}$&$0.83\pm0.03$&$-0.68\pm0.03$&$0.24\pm0.04$&$1.67\pm0.07$&$-3.53\pm0.32$\\
$\Xi_c^0 \rightarrow \Sigma^0 \eta$&$0.12\pm0.05$&$-0.02\pm0.71$&$0.25\pm0.18$&$2.18\pm0.61$&$4.68\pm1.11$&$\Xi_c^0 \rightarrow p \pi^{-}$&$0.11\pm0.01$&$-0.28\pm0.22$&$0.21\pm0.01$&$0.13\pm0.11$&$-2.15\pm0.65$\\
$\Xi_c^0 \rightarrow \Sigma^0 \eta'$&$0.20\pm0.04$&$-0.31\pm0.11$&$0.70\pm0.26$&$3.73\pm0.71$&$1.90\pm0.10$&$\Xi_c^0 \rightarrow n \eta$&$0.61\pm0.10$&$-0.40\pm0.07$&$0.20\pm0.06$&$1.14\pm0.15$&$2.17\pm5.08$\\
$\Xi_c^0 \rightarrow \Sigma^{-} \pi^{+}$&$0.48\pm0.02$&$-0.65\pm0.03$&$0.52\pm0.09$&$3.62\pm0.15$&$-3.53\pm0.32$&$\Xi_c^0 \rightarrow n \eta'$&$0.33\pm0.06$&$-0.45\pm0.17$&$0.20\pm0.07$&$1.06\pm0.17$&$-4.13\pm0.12$\\
$\Xi_c^0 \rightarrow \Xi^0 K_{S / L}$&$0.16\pm0.03$&$-0.08\pm0.31$&$0.47\pm0.19$&$2.48\pm0.34$&$-1.66\pm0.33$&\\
\hline
\hline
\end{tabular}
}
\end{table*}


\vskip 0.2cm
\noindent{\it  4. Conclusion---} 
\label{sec:dis}
We have performed a global fit to the experimental data of 2-body charmed baryon decays based on the topological diagrammatic approach  and taken
into account the phase shifts between $S$- and $P$-wave amplitudes. 
The measured branching fractions and decay asymmetries are well accommodated in TDA except for two modes, in particular, the predicted $\B(\Xi_c^0\to \Xi^-\pi^+)=(2.83\pm0.10)\%$ is larger than its current value.  The fit results for the branching fraction, decay asymmetry and the phase shift $\delta_P-\delta_S$ for $\Lambda_c^+\to \Xi^0K^+$ are consistent with the BESIII measurements. 

For yet-to-be-measured modes, we have presented the fitting magnitudes of $S$- and $P$-wave amplitudes and their phase shifts which can be tested in the near future. 

\vspace{0.5cm}
\begin{acknowledgments}
		We would like to thank Pei-Rong Li and Chia-Wei Liu for valuable discussions.
		This research was supported in part by the Ministry of Science and Technology of R.O.C. under Grant No. MOST-112-2112-M-001-026 and 
		the National Natural Science Foundation of China
		under Grant No. U1932104.

\end{acknowledgments}




\end{document}